\documentclass[twocolumn,showpacs,preprintnumbers,amsmath,amssymb]{revtex4}

\usepackage{graphicx}
\usepackage{dcolumn}
\usepackage{bm}

\begin{document}

\preprint{}
\input{epsf.tex}

\title{Hybrid Quantum System of a Nanofiber Mode Coupled to Two Chains of Optically Trapped Atoms}
\author{Hashem Zoubi, and Helmut Ritsch}

\affiliation{Institut f\"{u}r Theoretische Physik, Universit\"{a}t Innsbruck, Technikerstrasse 25, A-6020 Innsbruck, Austria}

\date{10 June, 2010}

\begin{abstract}
A tapered optical nanofiber simultaneously used to trap and optically interface of cold atoms through evanescent fields constitutes a new and well controllable hybrid quantum system. The atoms are trapped in two parallel 1D optical lattices generated by suitable far blue and red detuned evanescent field modes very close to opposite sides of the nanofiber surface. Collective electronic excitations (excitons) of each of the optical lattices are resonantly coupled to the second lattice forming symmetric and antisymmetric common excitons.  In contrast to the inverse cube dependence of the individual atomic dipole-dipole interaction, we analytically find an exponentially decaying coupling strength with distance between the lattices. The resulting symmetric (bright) excitons strongly interact with the resonant nanofiber photons to form fiber polaritons, which can be observed through linear optical spectra. For large enough wave vectors the polariton decay rate to free space is strongly reduced, which should render this system ideal for the realization of long range quantum communication between atomic ensembles.
\end{abstract}


\maketitle

\section{Introduction}

Effects related to light-matter interactions continue to be of big interests in the main present researches, due to their importance for fundamental physics and applications. Manipulation and trapping of atoms are few of the sequences of such interactions \cite{Metcalf}, and which serve as a framework for the development of quantum information technologies, e.g. the implementation of quantum memories, optical communications, and quantum computations \cite{Zeilinger}. Furthermore, trapping and guiding of atoms are useful tools for atom optics and atom interferometry \cite{Meystre}. Historically the system that take must attention was cavity quantum electrodynamics, which offers a powerful tool for controlling atom-photon interactions \cite{Haroche}, and even achieving the trapping of a single atom in a cavity \cite{Kimble, Rempe}. Recently, a Bose-Einstein condensate of an ultracold atom gas is fabricated between optical cavity mirrors, and were the strong electric dipole coupling regime achieved \cite{Esslinger, Reichel}. 

Different directions of researches in the past few years have been opened for the light-matter coupling which rest on optical fibers. In hollow core fiber the atoms are funneled into a capillary in the center of the fiber and they couple to the guided fiber photons, where this system can guide and confine both atoms and photons \cite{Ketterle,Vuletic,Bhagwat}. Other setup which is of importance for our present work is tapered optical fibers with a nanofiber waist, where here the atoms are outside the fiber and couple to the evanescent field surrounding the fiber \cite{Rauschenbeutel,Nayak}. This system combining both atomic and solid state devices and can be considered as hybrid quantum system, where trapping and optically interfacing of atoms can be simultaneously achieved. Moreover, the strong evanescent field around the nanofiber is efficiently used for trapping, detecting and manipulating of a single atom \cite{Hakuta}.

The guided modes of ultrathin optical fibers, with diameters smaller than the wavelength of the guided light, exhibit strong transverse confinement and pronounced evanescent field. Interference of two color evanescent fields surrounding an optical nanofiber give rise to an array of optical microtraps \cite{Sague}. The system realized recently for cesium atoms interacting with a multicolor evanescent field surrounding an optical nanofiber \cite{Vetsch}, where the atoms are localized in a one dimensional optical lattice parallel to the nanofiber, which shown to be efficiently interrogated with a resonant light field sent through the nanofiber. Moreover, conventional optical lattices are formed of counter propagating laser beams to get standing wave in free space, and then the ultracold atoms are loaded \cite{Bloch}. This system well described by Bose-Hubbard model which predicts the quantum phase transition from the superfluid into the Mott insulator phase with a fixed number of atoms per site \cite{Jaksch}.

In the present paper we investigate a simplified model related to the setup presented in \cite{Vetsch}. The hybrid quantum system build of a tapered optical nanofiber and two optical lattices. The one dimensional optical lattices formed by off resonance nanofiber modes appear parallel to each other localized at two opposite sides of the nanofiber. We consider the case of a single two-level atom per lattice site. The system is excited and probed by photons in a fiber mode resonant with the atomic transition. Due to resonant dipole-dipole interactions the electronic excitation dynamics for each chain is dominated by the formation of collective electronic excitations (excitons) incorporating the optical lattice translational symmetry \cite{ZoubiA,ZoubiB,ZoubiC}. Naturally excitons in the two chains interact directly and via the resonant fiber mode. In the limit of long wave length excitons and for a distance of the two chains larger than the lattice constant, the interaction among the two optical lattices can be calculated explicitly and strongly deviates from single particle dipole-dipole coupling. Eventually interlattice interactions give rise to dark and bright excitons which mix with the nanofiber photons in the strong coupling regime to form fiber polaritons \cite{ZoubiD}.

The paper is organized as follows. In section 2 we derive the exciton dispersions in two parallel optical lattices, and the exciton damping rate is presented. The excitons coupling to nanofiber photons is obtained in section 3. The strong coupling regime is treated in section 4 to give fiber polaritons. The transmission spectrum is calculated in section 5, and a summary appears in section 6. The appendix includes rigorous calculations of the itnra and inter lattices exciton dynamical matrices.

\section{Excitons in two parallel 1D optical lattices}

We consider two identical parallel one-dimensional optical lattices with one atom per site, where the lattice constant is $a$ and they are separated by a distance $d$. The optical lattice is located at a distance $b$ from the fiber surface, where the two optical lattices are at two opposite sides of the fiber, as seen in figure (1), which is related to the setup in \cite{Vetsch}. The total Hamiltonian of the hybrid quantum system is $H=H_{ex}+H_{ph}+H_{in}$, where $H_{ex}$ is the electronic excitations Hamiltonian in the two optical lattices, $H_{ph}$ is the fiber photons Hamiltonian, and $H_{in}$ is the coupling Hamiltonian between the fiber photons and the electronic excitations in the two optical lattices. First, we concentrate in the dynamics of electronic excitations in the two optical lattices. We examine the formation of hybrid collective electronic excitations (excitons) among the two lattices. An electronic excitation delocalizes in a lattice due to resonant dipole-dipole interactions and form an exciton which is a coherent state with wave number $k$ \cite{ZoubiA,ZoubiB}. Then we check if such an exciton can transfer between the two lattices due to resonant dipole-dipole interactions, and if it can coherently oscillate between the two lattices to form a collective electronic excitation in the whole system of the two optical lattices.

\begin{figure}[h!]
\centerline{\epsfxsize=8cm \epsfbox{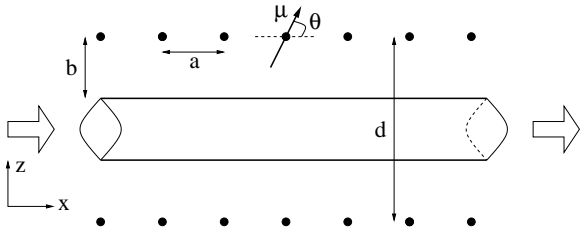}}
\caption{Two One-dimensional optical lattices with one atom per site, which are located parallel to a fiber at two opposite sides. An incident field is sent from the left side and the transmitted field is observed at the right side. The lattice constant is $a$ and the distance between the lattices is $d$. Here we assume $d\gg a$. The transition dipole $\mu$ which makes an angle $\theta$ with the lattice direction is also seen.}
\end{figure}

The electronic excitation Hamiltonian in the two optical lattices  is
\begin{equation}
H_{ex}=\sum_{n,\alpha}E_A\ B_{n\alpha}^{\dagger}B_{n\alpha}+\sum_{nm,\alpha\beta}J_{nm}^{\alpha\beta}\ B_{n\alpha}^{\dagger}B_{m\beta}.
\end{equation}
Here $(n,m)$ run over all the lattice sites, and $(\alpha,\beta)$ stands for the two lattices $(1,2)$. The atomic transition energy is $E_A=\hbar\omega_A$ for two level atoms, which is the same at the two lattices. $B_{n\alpha}^{\dagger},\ B_{n\alpha}$ are the creation and annihilation operators of an electronic excitation at site $n$ in lattice $\alpha$, respectively. We consider the case of a single electronic excitation or low excitation density, then the operators obey bosonic commutation relations. The coupling parameter $J_{nm}^{\alpha\beta}$ is for resonant dipole-dipole interactions, and give rise to excitation transfer among two sites. The transfer term include transfer among atoms in the same lattice, $J_{nm}^{11}$ and $J_{nm}^{22}$, and transfer among atoms at different lattices, $J_{nm}^{12}$ and $J_{nm}^{21}$.

The Hamiltonian can be diagonalized in the lattice sites in using the transformation
\begin{equation}\label{TRANS}
B_{n\alpha}=\frac{1}{\sqrt{N}}\sum_{k}e^{ikx_n^{\alpha}}B_{k\alpha},
\end{equation}
where $N$ is the number of lattice sites, which is taken to be large number, and $k$ is the wave number which takes the discrete values $k=\frac{2\pi}{Na}p$, with $(p=0,\pm 1,\pm 2,\cdots,\pm N/2)$, in using periodic boundary condition. Here $x_n^{\alpha}$ is the position of site $n$ in lattice $\alpha$. The new states represent collective electronic excitations which are called excitons \cite{ZoubiA}, and they are waves that propagate to the left and the right of the lattice with wave number $k$ which is a good quantum number. Now the Hamiltonian reads
\begin{equation}
H_{ex}=\sum_{k,\alpha}E_A\ B_{k\alpha}^{\dagger}B_{k\alpha}+\sum_{k,\alpha\beta}J^{\alpha\beta}(k)\ B_{k\alpha}^{\dagger}B_{k\beta},
\end{equation}
which is diagonal in $k$. We defined the exciton dynamical matrix by
\begin{equation}
J^{\alpha\beta}(k)=\sum_{L}e^{ikL}J^{\alpha\beta}(L),
\end{equation}
and we used $J_{nm}^{\alpha\beta}=J^{\alpha\beta}(L)$, with $L=x_m^{\beta}-x_n^{\alpha}$, where the dipole-dipole interaction is a function of the distance between the two atoms.

Next we diagonalize the above Hamiltonian relative to the two lattice indexes, $(\alpha,\beta)$, by applying the transformation
\begin{equation}\label{Symm}
B_{k\nu}=\frac{B_{k1}\pm B_{k2}}{\sqrt{2}}.
\end{equation}
The new states represent entangled states between excitons from the two lattices, which are symmetric and antisymmetric states. The symmetric state is denoted by $(\nu=s)$ and takes the plus sign $(+)$, the antisymmetric state is denoted by $(\nu=a)$ and takes the minus sign $(-)$. The Hamiltonian casts into the diagonal form
\begin{equation}
H_{ex}=\sum_{k,\nu}E_{ex}^{\nu}(k)\ B_{k\nu}^{\dagger}B_{k\nu},
\end{equation}
where the eigenenergies are
\begin{eqnarray}
E_{ex}^s(k)&=&E_A+J(k)+J'(k), \nonumber \\
E_{ex}^a(k)&=&E_A+J(k)-J'(k).
\end{eqnarray}
For identical lattices, inside the same lattice we have
\begin{equation}
J(k)=J^{11}(k)=J^{22}(k),
\end{equation}
and among different lattices we have
\begin{equation}
J'(k)=J^{12}(k)=J^{21}(k).
\end{equation}
Our main task now is to calculate the two exciton dynamical matrices, $J(k)$ and $J'(k)$.

\subsection{Exciton Dispersion}

The resonant dipole-dipole interaction between two atoms which are separated by a distance ${\bf R}$, and of transition dipole $\mbox{\boldmath$\mu$}$, is defined by
\begin{equation}\label{RDD}
J(\vec{R})=\frac{1}{4\pi\epsilon_0}\frac{|{\bf R}|^2|\mbox{\boldmath$\mu$}|^2-3(\mbox{\boldmath$\mu$}\cdot{\bf R})^2}{|{\bf R}|^5}.
\end{equation}
Such interaction holds in the limit of $|{\bf R}|<\lambda_A$, where $E_A=hc/\lambda_A$, otherwise one needs to include radiative corrections. For the dipole moment we take the general case of $\mbox{\boldmath$\mu$}=(\mu_x,\mu_y,\mu_z)$. The two optical lattices are parallel to the $x$ axis, one with $z_1=0$ and sites at $(la,0,0)$, and the other lattice with $z_2=d$ and sites at $(la,0,d)$, where $l$ is an integer that runs over the whole lattice sites. See figure (1).

In appendix A.1 we calculate the exciton dynamical matrix for interactions inside the same lattice, $J(k)$. In the limit of long wavelength excitons, that is $ka\ll 1$, we get
\begin{equation}
J(k)\approx \frac{\zeta(3)}{2\pi\epsilon_0}\frac{\mu_y^2+\mu_z^2-2\mu_x^2}{a^3}.
\end{equation}
Here the $y$ and $z$ components, which are normal to the lattice direction, are repulsive; and the $x$ component, which is parallel to the lattice direction, is attractive. Note that using the nearest neighbor interaction in this limit gives $J=\frac{1}{2\pi\epsilon_0}\frac{\mu_y^2+\mu_z^2-2\mu_x^2}{a^3}$.

Appendix A.2 includes the calculation for the exciton dynamical matrix for interaction among the two optical lattices, $J'(k)$. The result for the limit of long wavelength excitons, that is $ka\ll 1$, and large distance between the two lattice $kd\gg 1$, namely $d\gg a$, is given by
\begin{equation}
J'(k)\approx\frac{\sqrt{2\pi}}{ad^2}\frac{(kd)^{3/2}}{4\pi\epsilon_0}\ e^{-kd}\left\{\mu_x^2-\mu_z^2+\frac{4}{3}\frac{\mu_y^2}{kd}\right\}.
\end{equation}
Here the $x$ and $y$ terms are repulsive, as they parallel among the different lattices, but as the $x$ component is parallel to the lattice direction and the $y$ component is normal to it, the contribution of the $x$ component is much larger than the $y$ one. The $z$ term is attractive, even though these components are normal to the lattice direction they are parallel for different lattices. Approximately the last term is much smaller than the first two, (or we can assume $\mu_y=0$), we get
\begin{equation}
J'(ka\rightarrow 0,kd\rightarrow\infty)\approx\frac{\sqrt{2\pi}}{ad^2}\frac{(kd)^{3/2}}{4\pi\epsilon_0}\ e^{-kd}\left\{\mu_x^2-\mu_z^2\right\}.
\end{equation}
The main result of the present derivation is that the interaction decay exponentially with the distance between the two lattices. In place of the inverse cube dependence of the interaction between two dipoles, the collective effect in each lattice gives coupling between two excitons which is decay exponentially with the distance.

The exciton dispersions, $(\nu=a,s)$, for $\mu_y=0$, are
\begin{eqnarray}
E_{ex}^{\nu}(k)&\approx& E_A+\frac{\zeta(3)}{2\pi\epsilon_0a^3}\left\{\mu_z^2-2\mu_x^2\right\} \nonumber \\
&\pm&\frac{\sqrt{2\pi}}{ad^2}\frac{(kd)^{3/2}}{4\pi\epsilon_0}\ e^{-kd}\left\{\mu_x^2-\mu_z^2\right\}.
\end{eqnarray}

For this case of a dipole in the $(x-z)$ plane, with $\mbox{\boldmath$\mu$}=(\mu_x,0,\mu_z)=\mu(\cos\theta,0,\sin\theta)$, where $\theta$ is the angle between the dipole and the lattice direction, as seen in figure (1), we get
\begin{eqnarray}
E_{ex}^{\nu}(k,\theta)&\approx& E_A+\frac{\zeta(3)\mu^2}{2\pi\epsilon_0a^3}\left\{1-3\cos^2\theta\right\} \nonumber \\
&\pm&\frac{\sqrt{2\pi}}{ad^2}\frac{(kd)^{3/2}}{4\pi\epsilon_0}\ e^{-kd}\mu^2\left\{2\cos^2\theta-1\right\}.
\end{eqnarray}
The intralattice exciton dynamical matrix $J(k)$ is much larger than the interlattices one $J'(k)$, in the present limit. The splitting between the two exciton dispersions is
\begin{equation}
\Delta(k,\theta)=2\frac{\sqrt{2\pi}}{ad^2}\frac{(kd)^{3/2}}{4\pi\epsilon_0}\ e^{-kd}\mu^2\left\{2\cos^2\theta-1\right\},
\end{equation}
which is the coupling parameter for the coherent oscillation of the exciton between the two lattices. Such oscillation is possible only if the splitting is larger than the exciton damping rate. The symmetric-antisymmetric splitting vanishes at $\theta=45^o$, where we get the energy $E(\theta =45^o)\approx E_A-\frac{\zeta(3)\mu^2}{4\pi\epsilon_0a^3}$. At $\theta\approx 54.7^o$ the on-lattice term vanishes, where $1-3\cos^2\theta=0$, and we get
\begin{eqnarray}
E_{ex}^{\nu}(k,\theta\approx 54.7^o)&\approx& E_A\mp\frac{\sqrt{2\pi}}{3ad^2}\frac{(kd)^{3/2}\mu^2}{4\pi\epsilon_0}\ e^{-kd}.
\end{eqnarray}

\subsection{Exciton damping rate}

In spite the existence of the nanofiber, the optical lattices are located in free space, (in the next section we consider the coupling to the fiber photons). Therefore, we present here the damping rate of the exciton into free space. For a single one dimensional optical lattice with one atom per site we calculated in \cite{ZoubiE} the radiative damping rate into free space. The exciton damping rate is given by
\begin{eqnarray}
\Gamma_{ex}(k,\theta)&=&\frac{\mu^2E_{ex}^2(k)}{4\epsilon_0a\hbar^3c^2}\left\{1+\cos^2\theta\right. \nonumber \\
&-&\left.\frac{(\hbar ck)^2}{E_{ex}^2(k)}\left(2\cos^2\theta-\sin^2\theta\right)\right\},
\end{eqnarray}
which is a function of $k$ and $\theta$. Here the exciton energy is $E_{ex}(k)=E_A+J(k)$, which is as the above result but in neglecting $J'(k)$. This result needs to be compared with the single atom the damping rate, which is $\Gamma_{A}=\frac{\mu^2E_A^3}{3\pi\epsilon_0\hbar^4c^3}$. As we discussed in more details in \cite{ZoubiE}, the exciton damping rate is found to be much larger than $\Gamma_{A}$ for small wave numbers, that is $ka\ll 1$, and the excitons can be considered as superradiant states. For larger wave numbers, the damping rate of excitons with polarizations orthogonal to the lattice increases, while for polarizations parallel to the lattice decreases and becomes zero at a critical wave number. As the spontaneously emitted photon has wave vector component parallel to the lattice equals to the exciton wave number, the critical wave number is obtained from the conservation energy condition $E_{ex}^s(k_c)=\hbar ck_c$. Beyond this critical wave number, excitons become metastable with zero radiative damping rate, and they can propagate in the lattice without radiative decay.

In the present system we have two optical lattices, and we showed above that their mutual coupling give rise to symmetric and antisymmetric excitons. Direct calculations yield that the antisymmetric excitons are dark with zero damping rate, that is $\Gamma_{ex}^a=0$, while the symmetric excitons are bright with twice the single exciton damping rate, that is $\Gamma_{ex}^s=2\Gamma_{ex}$. Note that according to the previous discussion, the symmetric exciton damping rate also become zero beyond the critical wave number $k_c$. The dark excitons can not be excited directly through the coupling to the fiber photons and needs other methods to excite them. As we explain in the next section, the symmetric excitons can be excited directly through the electric dipole interaction with the fiber photons, where this coupling appears also for excitons with wave number beyond the critical wave number. This regime is useful for long range transmission through the system, and the only source of damping will be the fiber photon decay.

\section{Fiber photon-exciton interactions}

Now we treat the fiber photons and their coupling to the two optical lattices. Rigorous treatment of tapered nanofiber modes is given in \cite{Rauschenbeutel}, but here we use a simplified picture. The one-dimensional optical fiber modes are given by the Hamiltonian
\begin{equation}
H_{ph}=\sum_k\hbar\omega_{ph}(k)\ a_k^{\dagger}a_k,
\end{equation}
where $a_k^{\dagger}$ and $a_k$ are the fiber photon creation and annihilation operators with wave number $k$, respectively. The photon dispersion is taken to be of the simplified form
\begin{equation}
\omega_{ph}(k)=\frac{c}{\sqrt{\epsilon}}\sqrt{k_0^2+k^2},
\end{equation}
where the wave number $k_0$ results of the fiber transverse confinement, and $\epsilon$ is the fiber average dielectric constant. The electric field operator outside the fiber is
\begin{equation}
\hat{\bf E}({\bf r})=i\sum_k\sqrt{\frac{\hbar\omega_{ph}(k)}{2\epsilon_0 V}}\ {\bf e}\ u(r)\left\{a_k\ e^{ikz}-a_k^{\dagger}\ e^{-ikz}\right\},
\end{equation}
where ${\bf e}$ is the photon linear polarization unit vector, $V$ is the normalization volume, and $u(r)$ is the mode function which includes the fiber field complexity. The fiber photons can decay into free space with a given damping rate fixed by the nanofiber quality.

The atomic transition dipole operator is $\hat{\mbox{\boldmath$\mu$}}=\mbox{\boldmath$\mu$}\sum_{n,\alpha}\left(B_{n\alpha}+B_{n\alpha}^{\dagger}\right)$. The matter-light coupling is given by the electric dipole interaction $H_{in}=-\hat{\mbox{\boldmath$\mu$}}\cdot\hat{\bf E}$, and in the rotating wave approximation for linear polarizations, we get
\begin{eqnarray}
H_{in}&=&-i\sum_{k,n,\alpha}\sqrt{\frac{\hbar\omega_{ph}(k)}{2\epsilon_0 V}}\left(\mbox{\boldmath$\mu$}\cdot{\bf e}\right)u(b) \nonumber \\
&\times&\left\{a_kB_{n\alpha}^{\dagger}\ e^{ikx_n}-a_k^{\dagger}B_{n\alpha}\ e^{-ikx_n}\right\},
\end{eqnarray}
where in using the inverse transformation of equation (\ref{TRANS}), we have
\begin{eqnarray}
H_{in}&=&-i\sum_{kl,n,\alpha}u(b)\sqrt{\frac{\hbar\omega_{ph}(k)\mu^2}{2\epsilon_0 VN}} \nonumber \\
&\times&\left\{a_kB_{l\alpha}^{\dagger}\ e^{i(k-l)x_n}-a_k^{\dagger}B_{l\alpha}\ e^{-i(k-l)x_n}\right\}.
\end{eqnarray}
In using $\frac{1}{N}\sum_ne^{i(k-l)x_n}=\delta_{kl}$, we obtain
\begin{equation}
H_{in}=-i\sum_{k,\alpha}u(b)\sqrt{\frac{\hbar\omega_{ph}(k)\mu^2N}{2\epsilon_0 V}}\left\{a_kB_{k\alpha}^{\dagger}-a_k^{\dagger}B_{k\alpha}\right\}.
\end{equation}
We represent the exciton operators in terms of symmetric and antisymmetric operators by using the inverse of the transformation (\ref{Symm}), to get
\begin{equation}
H_{in}=-i\sum_{k}u(b)\sqrt{\frac{\hbar\omega_{ph}(k)\mu^2N}{\epsilon_0 V}}\left\{a_kB_{ks}^{\dagger}-a_k^{\dagger}B_{ks}\right\}.
\end{equation}
It is clear that only the symmetric excitons are coupled to the fiber photons, where the antisymmetric ones are dark and decouple to the photons.

In defining the coupling parameter
\begin{equation}
\hbar f_k=-iu(b)\sqrt{\frac{\hbar\omega_{ph}(k)\mu^2}{\epsilon_0 Sa}},
\end{equation}
where $S$ is the mode cross area, then we can write
\begin{equation}
H_{in}=\sum_{k}\left\{\hbar f_k\ a_kB_{ks}^{\dagger}+\hbar f_k^{\ast}\ a_k^{\dagger}B_{ks}\right\}.
\end{equation}
The photon dispersion and the exciton-photon coupling is taken to be $\theta$ independent.

\section{Fiber polaritons}

The total Hamiltonian is given by
\begin{eqnarray}
H&=&\sum_{k}\hbar\left\{\omega_{ex}^a(k,\theta)\ B_{ka}^{\dagger}B_{ka}+\omega_{ex}^s(k,\theta)\ B_{ks}^{\dagger}B_{ks}\right. \nonumber \\
&+&\left.\omega_{ph}(k)\ a_k^{\dagger}a_k+f_k\ a_kB_{ks}^{\dagger}+f_k^{\ast}\ a_k^{\dagger}B_{ks}\right\},
\end{eqnarray}
where due to translational symmetry along the lattice and fiber axis, the Hamiltonian is separated for each $k$.

In the strong coupling regime, where the coupling is larger than both the exciton and photon line width, we define polaritons by diagonalizing the Hamiltonian \cite{ZoubiA,ZoubiD}, to get the total Hamiltonian by
\begin{equation}
H=\sum_{kr}\hbar\omega^r_{pol}(k,\theta)\ A_k^{r\dagger}A^r_k+\sum_{k}\hbar\omega_{ex}^a(k,\theta)\ B_{ka}^{\dagger}B_{ka},
\end{equation}
with the polariton dispersions
\begin{equation}
\omega_{pol}^{\pm}(k,\theta)=\frac{\omega_{ph}(k)+\omega_{ex}^s(k,\theta)}{2}\pm \Delta(k,\theta),
\end{equation}
where
\begin{equation}
\Delta(k,\theta)=\sqrt{\delta^2(k,\theta)+|f_k|^2},
\end{equation}
with the detuning
\begin{equation}
\delta(k,\theta)=\frac{\omega_{ph}(k)-\omega_{ex}^s(k,\theta)}{2}.
\end{equation}
The polariton operators are a coherent superposition of symmetric excitons and photons, where
\begin{equation}
A_k^{\pm}=X^{\pm}(k,\theta)\ B_{ks}+Y^{\pm}(k,\theta)\ a_k,
\end{equation}
and where the amplitudes are given by
\begin{eqnarray}
X^{\pm}(k,\theta)&=&\pm\sqrt{\frac{\Delta(k,\theta)\mp\delta(k,\theta)}{2\Delta(k,\theta)}}, \nonumber \\
Y^{\pm}(k,\theta)&=&\frac{f_k}{\sqrt{2\Delta(k,\theta)(\Delta(k,\theta)\mp\delta(k,\theta))}}.
\end{eqnarray}

Here we present the results for a system with the following parameters: the lattice constant is $a=1000\ \AA$, the transition energy is $E_A=1\ eV$, the transition dipole is $\mu=1\ e\AA$ of angle $\theta=90^o$ with the lattice axis. The distance between the two optical lattices is taken to be $d=10a$. The fiber dielectric constant is $\epsilon=3$, with the mode cross area of $S=4\pi a^2$, and the mode function at the lattice position is taken to be $u(b)=0.1$. The cavity mode energy at $k=0$ is taken to be in resonance with the free atom transition energy, that is $E_{ph}(k=0)=E_A$, then $k_0=E_A\sqrt{\epsilon}/\hbar c$. The polariton dispersions are plotted in figure (2) relative to the free atom transition, and the symmetric exciton and photon weights are plotted in figure (3). The exciton-photon intersection point is taken here to be at $k=0$, where the polaritons split by the Rabi splitting, and they are half exciton and half photon. For large $k$ the upper branch becomes photonic, and the lower becomes excitonic.

\begin{figure}[h!]
\centerline{\epsfxsize=7cm \epsfbox{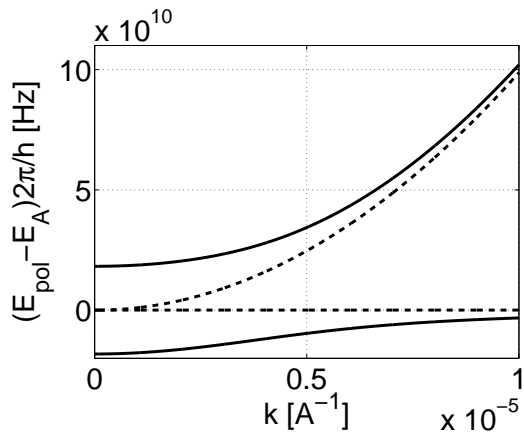}}
\caption{The full lines are the upper and lower polariton angular frequencies relative to the free atom transition, $\omega_{pol}-\omega_A$ vs. wave number $k$. The dashed line is the symmetric exciton dispersion and the dashed parabola is for the fiber photon dispersion.}
\end{figure}

\begin{figure}[h!]
\centerline{\epsfxsize=7cm \epsfbox{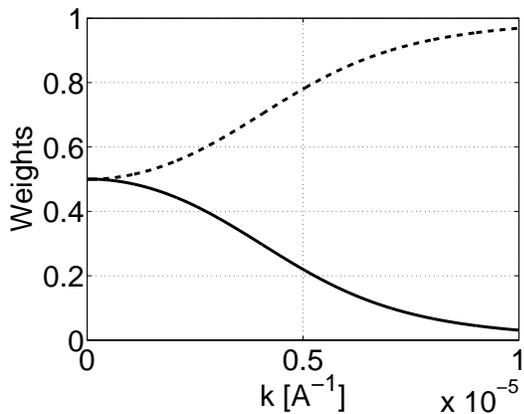}}
\caption{The excitonic and photonic weights vs. $k$. For the lower polariton branch the dashed line is for the excitonic weight, and the full line for the photonic weight. For the upper polariton branch the dashed line is for the photonic weight, and the full line for the excitonic weight.}
\end{figure}

\section{Fiber transmission spectrum}

To get the linear optical spectra \cite{ZoubiA}, we consider an incident field from the far left side of the fiber, and we calculate the transmission $T(k,\theta)$ and reflection $R(k,\theta)$ spectra, with the absorption spectrum $A(k,\theta)$. The coupling of the fiber field to the external field at the two far edges of the fiber is included in the parameter $\gamma$, and the fiber photon damping rate is taken to be $\Gamma_{ph}$ which is assumed to be $k$ independent. The symmetric exciton damping rate is included here phenomenologically, and is taken to be $\Gamma_{ex}^s(k,\theta)$ as discussed in section 2.A. The polariton damping rate is taken to be
\begin{equation}
\Gamma_{pol}^{\pm}(k,\theta)=\Gamma_{ex}^s(k,\theta)\ |X^{\pm}(k,\theta)|^2+\Gamma_{ph}\ |Y^{\pm}(k,\theta)|^2.
\end{equation}

The linear optical spectra are calculated and given by
\begin{equation}
R(\omega,k,\theta)=\frac{1}{|1+i\gamma\ \Lambda(\omega,k,\theta)|^2},
\end{equation}
and
\begin{equation}
T(\omega,k,\theta)=\frac{\gamma^2\ |\Lambda(\omega,k,\theta)|^2}{|1+i\gamma\ \Lambda(\omega,k,\theta)|^2},
\end{equation}
with
\begin{equation}
R(\omega,k,\theta)+T(\omega,k,\theta)+A(\omega,k,\theta)=1,
\end{equation}
where
\begin{equation}
\Lambda(\omega,k,\theta)=\sum_r\frac{|Y^r(k,\theta)|^2}{\omega-\bar{\omega}^r_{pol}(k,\theta)},
\end{equation}
where $\bar{\omega}^r_{pol}(k,\theta)=\omega_{pol}^{r}(k,\theta)-i\Gamma_{pol}^{\pm}(k,\theta)$.

In the linear spectra plots we use for the photon a small damping rate of $\hbar\Gamma_{ph}=10^{-10}\ eV$, and the fiber edge coupling parameter is taken to have a bandwidth of $\hbar\gamma=10^{-6}\ eV$. In figure (4) we plot the transmission spectra for a wave number of $k=10^{-6}\ \AA^{-1}$, where we get two transmission peaks correspond to the two polariton branches which are separated by the Rabi splitting. Now we take an incident field of a wide bandwidth of $\hbar\gamma=10^{-4}\ eV$. In figure (5) we plot the transmission spectra for a wave number of $k=10^{-6}\ \AA^{-1}$, where here the band gab between the two peaks becomes a dip in the transmission spectrum. Figure (6) is a close look at the minimum which show a small shift relative to the free atom transition, and that results of the dipole-dipole interactions. Here the symmetric exciton line width is $\hbar\Gamma_{ex}^s=2.32\times 10^{-8}\ eV$, which is larger than the free atom line width of $\hbar\Gamma_A=2.5\times 10^{-9}\ eV$. In figure (7) we plot the transmission spectra for wave number of $k=10^{-5}\ \AA^{-1}$, and in figures (8-9) for $k=5\times 10^{-5}\ \AA^{-1}$, where in figure (8) the plot is around the upper polariton branch, and in figure (9) is around the lower polariton branch which is also seen in figure (8) as a small peak around zero. It is clear that for large wave numbers at the lower branch we get small transmission, where for large $k$ the excitons and photons with the same wave number have large detuning, namely they are off resonance. At the upper branch we get large transmission. As we mentioned previously, the upper branch becomes photonic and the lower one excitonic.

\begin{figure}[h!]
\centerline{\epsfxsize=7cm \epsfbox{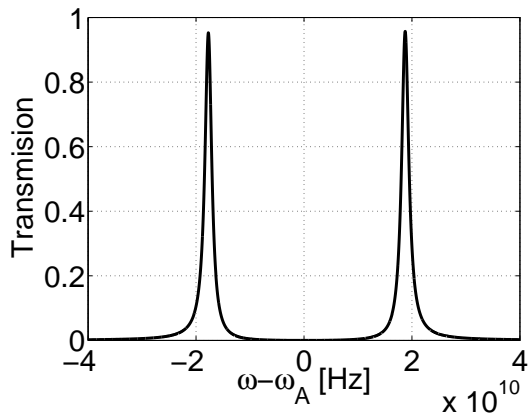}}
\caption{The transmission spectrum, for $k=10^{-6}\ \AA^{-1}$, and $\hbar\gamma=10^{-6}\ eV$. The two peaks correspond to the two polariton branches.}
\end{figure}

\begin{figure}[h!]
\centerline{\epsfxsize=7cm \epsfbox{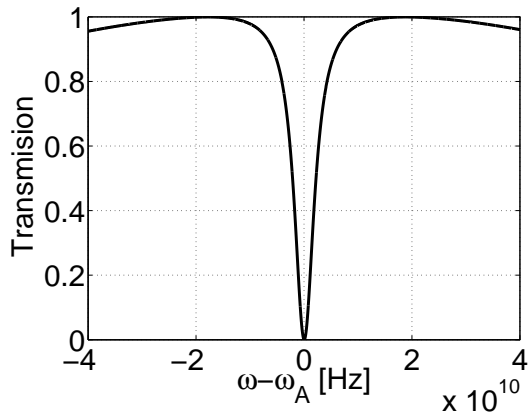}}
\caption{The transmission spectrum, for $k=10^{-6}\ \AA^{-1}$, and $\hbar\gamma=10^{-4}\ eV$.}
\end{figure}

\begin{figure}[h!]
\centerline{\epsfxsize=7cm \epsfbox{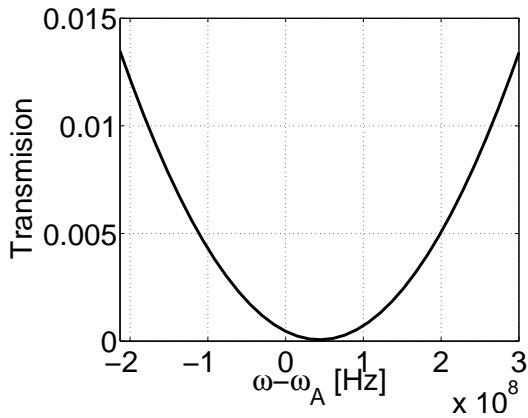}}
\caption{The transmission spectrum, for $k=10^{-6}\ \AA^{-1}$, and $\hbar\gamma=10^{-4}\ eV$. A close look around the minimum.}
\end{figure}

\begin{figure}[h!]
\centerline{\epsfxsize=7cm \epsfbox{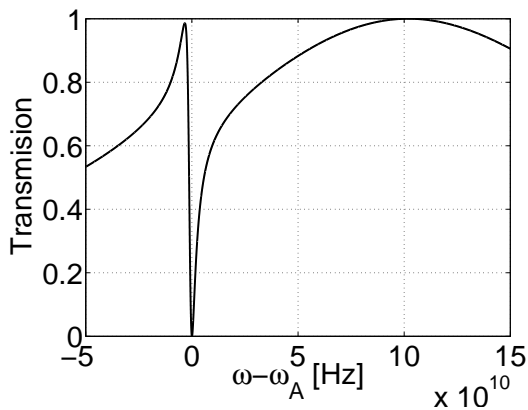}}
\caption{The transmission spectrum, for $k=10^{-5}\ \AA^{-1}$.}
\end{figure}

\begin{figure}[h!]
\centerline{\epsfxsize=7cm \epsfbox{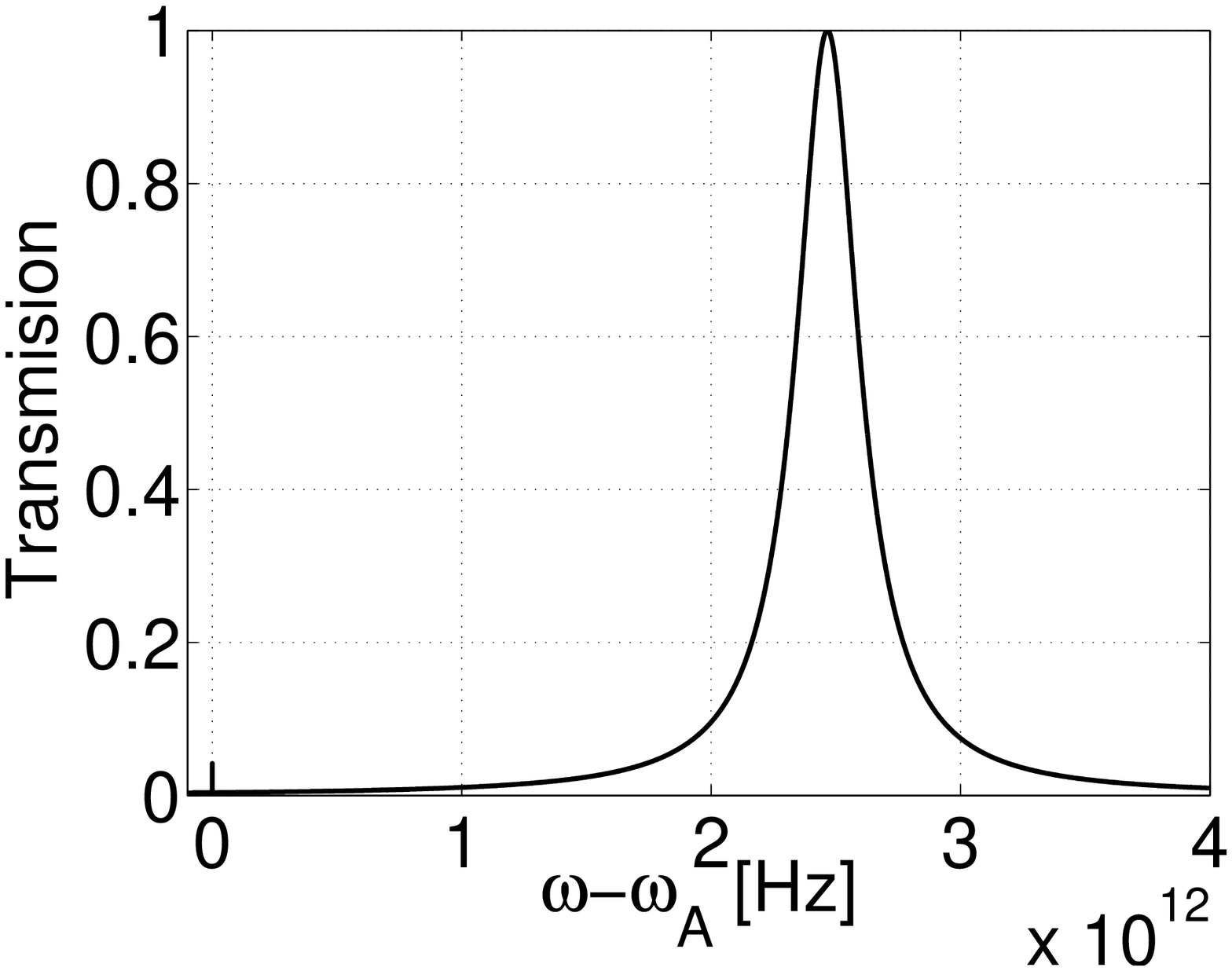}}
\caption{The transmission spectrum, for $k=5\times 10^{-5}\ \AA^{-1}$, around the upper polariton branch.}
\end{figure}

\begin{figure}[h!]
\centerline{\epsfxsize=7cm \epsfbox{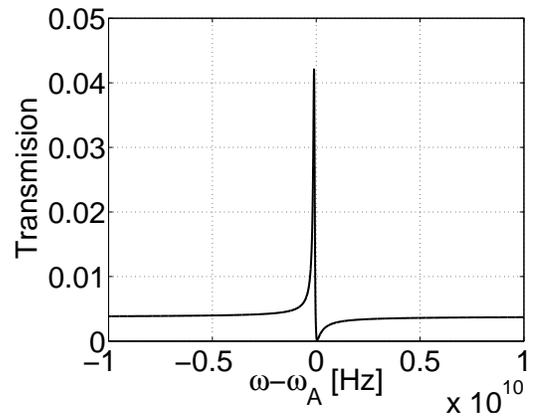}}
\caption{The transmission spectrum, for $k=5\times 10^{-5}\ \AA^{-1}$, around the lower polariton branch.}
\end{figure}

\section{Summary}

We studied a recently demonstrated integrated hybrid atom-photon quantum system based on a tapered nanofiber and cold atoms. The nanofiber is used to implement two arrays of nanotraps for the atoms using off-resonant photons as well as to spatially confine resonant photons efficient to ensure strong single photon coupling to the two atomic ensembles. Interestingly the effective coupling parameter of excitons in the two chains decays exponentially with their distance but it is still sufficient to split the energies of symmetric and antisymmetric excitons. Only the symmetric excitons are bright and coherently mix with the nanofiber photons to form  fiber polaritons, which can be seen in the transmission spectrum of the nanofiber.

The present system allow one to investigate and to deeply understand one dimensional collective electronic excitations (excitons), where it is possible to excite, store, and readout information in the form of coherent electronic excitations. As was shown in our previous work \cite{ZoubiE}, the damping rate of small wave number excitons in one dimensional lattice is much larger than a single atom excitation, and decay fast into free space. But beyond a critical wave number it is shown that one dimensional excitons became metastable with infinite radiative life time. In this regime, excitons can not be excited by free space photons, but can be easily excited by coupling to fiber photons. These excitons can serve for information storage devices and for communications through energy transfer. In the present paper the discussion is limited for the linear regime. Nonlinear processes induced by electronic excitation saturation effects give rise for more interesting physics, which is our next task. At this point we have neglected the interaction of the atoms via light forces induced by multiple scattering of photons \cite{Asboth}, which can be significantly enhanced through the transverse mode confinement and the fact that the atoms form Bragg like structures. This could lead to collective oscillations and instabilities of the lattices but also can be used to engineer long distance motional couplings between ensembles. 

\begin{acknowledgments}
The work was supported by the Austrian Science Funds (FWF) via the project (P21101), and the Eranet NanoSci-E+ project NOI (I269-N16).
\end{acknowledgments}

\appendix

\section{Exciton dynamical matrix calculations}

In this appendix we calculate the exciton dynamical matrix for the two cases of intralattice interactions, $J(k)$, and interactions among interlattices, $J'(k)$.

\subsection{Intralattice interactions}

For interactions inside the same lattice, from equation (\ref{RDD}), we have the resonant dipole-dipole interaction
\begin{equation}
J(l)=\frac{1}{4\pi\epsilon_0}\frac{\mu_y^2+\mu_z^2-2\mu_x^2}{a^3l^3},
\end{equation}
and in momentum space reads
\begin{eqnarray}
J(k)&=&\frac{1}{4\pi\epsilon_0}\frac{\mu_y^2+\mu_z^2-2\mu_x^2}{a^3}\sum_{l=-\infty}^{+\infty\ \prime}\frac{e^{ikal}}{l^3} \nonumber \\
&=&\frac{1}{4\pi\epsilon_0}\frac{\mu_y^2+\mu_z^2-2\mu_x^2}{a^3}\sum_{l=1}^{\infty}\frac{2\cos(kal)}{l^3},
\end{eqnarray}
the prime indicate that $l$ runs from $-\infty$ to $+\infty$ in excluding $l=0$ which is for self interactions. In the limit of long wavelength excitons, or small wave number, that is $ka\ll 1$, we get
\begin{eqnarray}
J(ka\rightarrow 0)&\simeq& \frac{1}{4\pi\epsilon_0}\frac{\mu_y^2+\mu_z^2-2\mu_x^2}{a^3}\sum_{l=1}^{\infty}\frac{2}{l^3} \nonumber \\
&=&\frac{1}{2\pi\epsilon_0}\zeta(3)\frac{\mu_y^2+\mu_z^2-2\mu_x^2}{a^3},
\end{eqnarray}
where the Riemann zeta function gives $\zeta(3)\approx 1.202$.

\subsection{Interlattices interactions}

In order to calculate the exciton dynamical matrix for interactions among the two lattices, we use the Ewald's method \cite{Born}, which is applied here for excitons in one dimensional optical lattices.

The resonant dipole-dipole interaction reads
\begin{equation}
J'(l)=\sum_{ij}\frac{\mu_i\mu_j}{4\pi\epsilon_0}\ D_{ij}(l),
\end{equation}
where $(i,j=x,y,z)$. The calculations give
\begin{eqnarray}
D_{xx}(l)&=&\frac{1}{(a^2l^2+d^2)^{3/2}}-\frac{3a^2l^2}{(a^2l^2+d^2)^{5/2}}, \nonumber \\
D_{yy}(l)&=&\frac{1}{(a^2l^2+d^2)^{3/2}}, \nonumber \\
D_{zz}(l)&=&\frac{1}{(a^2l^2+d^2)^{3/2}}-\frac{3d^2}{(a^2l^2+d^2)^{5/2}}, \nonumber \\
D_{xz}(l)&=&D_{zx}(l)=-\frac{3adl}{(a^2l^2+d^2)^{5/2}}, \nonumber \\
D_{xy}(l)&=&D_{yx}(l)=D_{yz}(l)=D_{zy}(l)=0.
\end{eqnarray}
We need to calculate the interactions in momentum space
\begin{equation}
J'(k)=\sum_{ij}\frac{\mu_i\mu_j}{4\pi\epsilon_0}\ D_{ij}(k),
\end{equation}
where
\begin{equation}
D_{ij}(k)=\sum_{l=-\infty}^{+\infty} e^{ikal}D_{ij}(l).
\end{equation}
We define the function
\begin{equation}
S(k)=\sum_{l=-\infty}^{+\infty}\frac{e^{ikal}}{(a^2l^2+d^2)^{5/2}},
\end{equation}
then the matrix elements of the exciton dynamical matrix can be derived from it by
\begin{eqnarray}
D_{xx}(k)&=&\left(2\frac{\partial^2}{\partial k^2}+d^2\right)S(k), \nonumber \\
D_{yy}(k)&=&\left(-\frac{\partial^2}{\partial k^2}+d^2\right)S(k), \nonumber \\
D_{zz}(k)&=&\left(-\frac{\partial^2}{\partial k^2}-2d^2\right)S(k), \nonumber \\
D_{xz}(k)&=&D_{zx}^{\ast}(k)=i3d\frac{\partial}{\partial k}S(k).
\end{eqnarray}

The function $S(k)$ can be written in a form of an integral with exponential decay, in place of the $1/R^5$ dependence, in using the identity
\begin{equation}
\frac{4}{3\sqrt{\pi}}\int_0^{\infty}dt\ t^{3/2}e^{-\alpha t}=\frac{1}{\alpha^{5/2}}.
\end{equation}
Then we arrive at
\begin{equation}
S(k)=\frac{4}{3\sqrt{\pi}}\int_0^{\infty}dt\ t^{3/2}e^{-d^2 t}\sum_{l=-\infty}^{+\infty}e^{-a^2l^2 t+ikal}.
\end{equation}
In using the relation
\begin{equation}
\sum_{l=-\infty}^{+\infty}e^{-a^2l^2 t+ikal}=\frac{\sqrt{\pi}}{a\sqrt{t}}\sum_{n=-\infty}^{+\infty}e^{-\frac{1}{a^2t}\left(n\pi+\frac{ka}{2}\right)^2}.
\end{equation}
We convert one summation into another. The first includes oscillations, and then decays slowly for large summation index $l$. While the second decays faster with increasing the summation index $n$. Such a change in the series found to be useful for long wavelength excitons, as we show in the following. In doing this, we obtain
\begin{equation}
S(k)=\frac{4}{3a}\sum_{n=-\infty}^{+\infty}\int_0^{\infty}dt\ t\ e^{-d^2 t}e^{-\frac{1}{a^2t}\left(n\pi+\frac{ka}{2}\right)^2}.
\end{equation}
The integral gives
\begin{equation}
\int_0^{\infty}dt\ t\ e^{-\frac{\beta}{t}-\gamma t}=2\frac{\beta}{\gamma}K_2(2\sqrt{\beta\gamma}),
\end{equation}
where $K_n(x)$ is the modified Bessel function of the second kind of order $n$. Finally we have
\begin{equation}
S(k)=\frac{4}{3a}\sum_{n=-\infty}^{+\infty}\frac{2\left(n\pi+\frac{ka}{2}\right)^2}{a^2d^2}K_2\left(\frac{2d\left(n\pi+\frac{ka}{2}\right)}{a}\right).
\end{equation}

We interest in the limit of long wavelength excitons, that is $ka\ll 1$. In this limit the dominant contribution comes from the term with $(n=0)$, and the other terms decay faster, hence we get
\begin{equation}
S(ka\rightarrow 0)\simeq\frac{2}{3}\frac{k^2}{ad^2}K_2(kd).
\end{equation}
We need to use the relation
\begin{equation}
\frac{\partial}{\partial x}K_n(x)=-\frac{1}{2}[K_{n-1}(x)+K_{n+1}(x)].
\end{equation}
Here we use the second main approximation, which is for large $d$ where $kd\gg 1$, and as $ka\ll 1$, we are in the limit of $d\gg a$. For the function $K_{n}(kd)$ we use the asymptotic expansion
\begin{equation}
K_{n}(kd\rightarrow\infty)\approx \sqrt{\frac{\pi}{2kd}}e^{-kd},
\end{equation}
which has the same behavior for different $n$. In this limit we get the matrix elements
\begin{eqnarray}
D_{xx}(ka\rightarrow 0,kd\rightarrow\infty)&\approx&\frac{\sqrt{2\pi}}{ad^2}(kd)^{3/2}\ e^{-kd}, \nonumber \\
D_{yy}(ka\rightarrow 0,kd\rightarrow\infty)&\approx&\frac{4}{3}\frac{\sqrt{2\pi}}{ad^2}(kd)^{1/2}\ e^{-kd}, \nonumber \\
D_{zz}(ka\rightarrow 0,kd\rightarrow\infty)&\approx&-\frac{\sqrt{2\pi}}{ad^2}(kd)^{3/2}\ e^{-kd}, \nonumber \\
D_{xz}(ka\rightarrow 0,kd\rightarrow\infty)&\approx&-2i\frac{\sqrt{2\pi}}{ad^2}(kd)^{3/2}\ e^{-kd}. \nonumber \\
\end{eqnarray}

Finally we obtain
\begin{eqnarray}
J'(ka\rightarrow 0,kd\rightarrow\infty)&\approx&\frac{\sqrt{2\pi}}{ad^2}\frac{(kd)^{3/2}}{4\pi\epsilon_0}\ e^{-kd} \nonumber \\
&\times&\left\{\mu_x^2-\mu_z^2+\frac{4}{3}\frac{\mu_y^2}{kd}\right\}.
\end{eqnarray}



\begin{thebibliography}{9}

\bibitem{Metcalf} H. J. Metcalf, and P. van der Straten, {\it Laser Cooling and Trapping}, (Springer, NY, 1999).

\bibitem{Zeilinger} D. Bouwmeester, A. K. Ekert, and A. Zeilinger, {\it The physics of Quantum Information}, (Springer, NY, 2000).

\bibitem{Meystre} P. Meystre, {\it Atom Optics}, (Springer, NY, 2001).

\bibitem{Haroche} S. Haroche, J. M. Raimond, {\it Exploring the Quantum: Atoms, Cavities and Photons}, (Oxford UP, UK, 2006).

\bibitem{Kimble} J. Ye, D. W. Vernooy, and H. J. Kimble, {\it Phys. Rev. Lett.} {\bf 83}, 4987 (1999).

\bibitem{Rempe} P. W. H. Pinkse, T. Fischer, P. Maunz, and G. Rempe, {\it Nature} {\bf 404}, 365 (2000).

\bibitem{Esslinger} F. Brennecke, T. Donner, S. Ritter, T. Bourdel, M. Köhl, and T. Esslinger, {\it Nature} {\bf 450}, 268 (2007).

\bibitem{Reichel} Y. Colombe, T. Steinmetz, G. Dubois, F. Linke, D. Hunger, and J. Reichel, {\it Nature} {\bf 450}, 272 (2007).

\bibitem{Verdu} J. Verdu, H. Zoubi, C. Koller, J. Majer, H. Ritsch, and J. Schmiedmayer, {\it Phys. Rev. Lett.} {\bf 103}, 043603 (2009).

\bibitem{Atac} A. Imamoglu, {\it Phys. Rev. Lett.} {\bf 102}, 083602 (2009).

\bibitem{Ketterle} C. A. Christensen, S. Will, M. Saba, G. B. Jo, Y. I. Shin, W. Ketterle, and D. Pritchard, {\it Phys. Rev. A} {\bf 78}, 033429 (2008).

\bibitem{Vuletic} M. Bajcsy, S. Hofferberth, V. Balic, T. Peyronel, M. Hafezi, A. S. Zibrov, V. Vuletic, and M. D. Lukin, {\it Phys. Rev. Lett.} {\bf 102}, 203902 (2009).

\bibitem{Bhagwat} A. R. Bhagwat, and A. L. Gaeta, {\it Opt. Express} {\bf 16}, 5035 (2008).

\bibitem{Rauschenbeutel} G. Sague, E. Vetsch, W. Alt, D. Meschede, and A. Rauschenbeutel, {\it Phys. Rev. Lett.} {\bf 99}, 163602 (2007).

\bibitem{Nayak} K. P. Nayak, P. N. Melentiev, M. Morinaga, F. L. Kien, V. I. Balykin, and K. Hakuta, {\it Opt. Express} {\bf 15}, 5431 (2007).

\bibitem{Hakuta} K. P. Nayak, and K. Hakuta, {\it New J. Phys.} {\bf 10}, 053003 (2008).

\bibitem{Sague} G. Sague, A. Baade, and A. Rauschenbeutel, {\it New J. Phys.} {\bf 10}, 113008 (2008).

\bibitem{Vetsch} E. Vetsch, D. Reitz, G. Sague, R. Schmidt, S. T. Dawkins, and A. Rauschenbeutel, {\it Phys. Rev. Lett.} {\bf 104}, 203603 (2010).

\bibitem{Bloch} M. Greiner, O. Mandel, T. Esslinger, T. W. Hansch, and I. Bloch, {\it Nature} {\bf 415}, 39 (2002).

\bibitem{Jaksch} D. Jaksch, C. Bruder, J. I. Cirac, C. W. Gardiner, and P. Zoller, {\it Phys. Rev. Lett.} {\bf 81}, 3108 (1998).

\bibitem{ZoubiA} H. Zoubi, and H. Ritsch, {\it Phys. Rev. A} {\bf 76}, 013817 (2007).

\bibitem{ZoubiB} H. Zoubi, and H. Ritsch, {\it Europhys. Lett.} {\bf 82}, 14001 (2008).

\bibitem{ZoubiC} H. Zoubi, and H. Ritsch, {\it New J. Phys.} {\bf 10}, 23001 (2008).

\bibitem{ZoubiD} H. Zoubi, and H. Ritsch, {\it Europhys. Lett.} {\bf 87}, 23001 (2009).

\bibitem{ZoubiE} H. Zoubi, and H. Ritsch, {\it Europhys. Lett.} {\bf 90}, 23001 (2010).

\bibitem{Asboth} J. K. Asboth, H. Ritsch, and P. Domokos, {\it Phys. Rev. Lett.} {\bf 98}, 203008 (2007).

\bibitem{Born} M. Born, and K. Huang, {\it Dynamical Theory of Crystal Lattices}, (Clarendon Press, Oxford,1954).

\end{thebibliography}
\end{document}